\documentclass[preprint2]{aastex}

\usepackage{epsfig}

\usepackage{amssymb}	%needed for "\lesssim" & "\gtrsim"

\newcommand{\OI}{\mbox{\rm [O\I]}}
\newcommand{\I}{\protect\small I \normalsize $\!\!$}
\newcommand{\II}{\protect\small II \normalsize $\!\!$}

\newcommand{\CII}{\mbox{\rm [C\II}]}
\newcommand{\HI}{\mbox{H\,{\sc i}}}

\newcommand{\Htwo}{\mbox{\rm H$_{2}$}}

\newcommand{\pcmcub}{\mbox{${\rm cm^{-3}}$}}
\newcommand{\ps}{\mbox{${\rm s^{-1}}$}}
\newcommand{\psr}{\mbox{${\rm sr^{-1}}$}}
\newcommand{\kmps}{\mbox{${\rm km\;s^{-1}}$}}

\newcommand{\pcmsq}{\mbox{${\rm cm^{-2}}$}}
\newcommand{\pwr}[2]{\mbox{$#1 \times 10^{#2}$}}

\newcommand{\Gzero}{$G_0$}

\newcommand{\lsim}{\mbox{$\mathrel{\vcenter{\hbox{\ooalign{\raise3pt\hbox{$<$}\crcr \lower3pt\hbox{$\sim$}}}}}$}}
\newcommand{\gsim}{\mbox{$\mathrel{\vcenter{\hbox{\ooalign{\raise3pt\hbox{$>$}\crcr \lower3pt\hbox{$\sim$}}}}}$}}

\slugcomment{Submitted to the Astrophysical Journal}

\shorttitle{\HI\ and CO Emission from PDRs}
\shortauthors{Allen, Heaton, \& Kaufman}

\received{2003 April 9}		% added by the ApJ editor

\begin{document}

%\title{THE PRODUCTION OF \HI\ IN PHOTODISSOCIATION REGIONS AND \\
%A COMPARISON WITH CO(1--0) EMISSION}
%
\title{The Production of \HI\ in Photodissociation Regions and \\
A Comparison with CO(1--0) Emission \\[0.4in]}
\author{Ronald J.~Allen}
\affil{Space Telescope Science Institute\\
3700 San Martin Drive, Baltimore, MD 21218;
\email{rjallen@stsci.edu}}
\author{Harold I.~Heaton}
\affil{Johns Hopkins University, Applied Physics Laboratory\\
11,100 Johns Hopkins Road, Laurel, MD 20723;
\email{Hal.Heaton@jhuapl.edu}}
\and
\author{Michael J.~Kaufman}
\affil{Department of Physics, San Jose State University\\
One Washington Square, San Jose, CA 95192-0106;
\email{kaufman@ism.arc.nasa.gov}}
\begin{abstract}
The gas at the surfaces of molecular clouds in galaxies is heated and
dissociated by photons from young stars both near and far.  \HI\ resulting
from the dissociation of molecular hydrogen \Htwo\ emits hyperfine line
emission at $21$ cm, and warmed CO emits dipole rotational lines such as the
$2.6$ mm line of CO(1--0).  We use previously developed models for
photodissociation regions (PDRs) to compute the intensities of these \HI\ and
CO(1--0) lines as a function of the total volume density $n$ in the cloud and
the far ultraviolet flux \Gzero\ incident upon it and present the results in
units familiar to observers.  The intensities of these two lines behave
differently with changing physical conditions in the PDR, and, taken together,
the two lines can provide a ground--based radio astronomy diagnostic for
determining $n$ and \Gzero\ separately in distant molecular clouds.  This
diagnostic is particularly useful in the range \Gzero\ $\lesssim 100$, 10
\pcmcub $\lesssim n \lesssim 10^5$ \pcmcub, which applies to a large fraction
of the volume of the interstellar medium in galaxies.  If the molecular cloud
is located near discrete sources of far--UV (FUV) emission, the PDR--generated
\HI\ and CO(1--0) emission on the cloud surface can be more easily identified,
appearing as layered ``blankets'' or ``blisters'' on the side of the cloud
nearest to the FUV source.  As an illustration, we consider the Galactic
object G216 -2.5, i.e.\ ``Maddalena's Cloud'', which has been previously
identified as a large PDR in the Galaxy.  We determine that this cloud has
$n\approx 200$ \pcmcub\ and \Gzero\ $\approx 0.8$, consistent with other data.
\end{abstract}


\section{Introduction}
\label{sec:intro}

The interstellar medium (ISM) in galaxies is excited, dissociated, and ionized
by far-ultraviolet (FUV) photons produced by young O and B stars.  Atomic gas
in the ISM recombines into molecular form mainly through the catalytic action
of dust grain surfaces.  In particular, hydrogen nuclei in the ISM cycle
repeatedly from the molecular (\Htwo) to the atomic (\HI) phase and back
again, at rates depending on the incident FUV flux \Gzero, the total volume
density $n$ of the gas, and the dust-to-gas ratio $\delta$.  Regions in the
ISM where the physics is dominated by FUV photons are called photodissociation
regions (PDRs).  The surfaces of giant molecular clouds (GMCs) are important
(and ubiquitous) examples of PDRs in galaxies.

The physics of PDRs has been explored in detail over the past $\sim\,30$ years
by many workers including D.\ J.\ Hollenbach, B.\ T.\ Draine, A.\ Dalgarno,
J.\ H.\ Black, A.\ G.\ G.\ M.\ Tielens, E.\ van Dishoeck, J.\ Le Bourlot, and
their students and collaborators.  One major focus of this work has been to
explain the $\approx 1$\% line-to-continuum ratios of the far-infrared lines
of \CII\ and \OI\ in Galactic sources observed from high-altitude aircraft and
balloon platforms.  With the advent of more sensitive space-based
observations, the models were extended to include the weak
rotational-vibrational spectrum of excited \Htwo\ observed on the active PDR
surfaces of GMCs that are exposed to relatively intense FUV fluxes from nearby
young stars.  An excellent review of the observational and theoretical state
of the field is given by \citet{hol99}.

PDR model computations have become very detailed and comprehensive, raising
the possibility that observations of multiple spectral lines including those
from trace molecules such as CO can be used to "invert" the models in order to
determine physical conditions in the ISM.  For example, a set of such models
has been computed by \cite{kau99} for use in the interpretation of
far-infrared and submillimeter spectra from Galactic and extragalactic
sources.  However, we wish to point out that, although there is qualitative
agreement on most of the physics, there are differences in the details of the
various model codes currently in use around the world as well as differences
in the numerical values of the parameters adopted.  These differences can lead
to disagreements in the values derived for physical conditions even for the
same observational data:  an example of such differences is illustrated in
\cite{li02}.  Efforts are underway to intercompare models from different
groups in a more systematic manner (E.\ van Dishoeck 2003, private
communication).  The results obtained from any one model computation (such as
the one we use here) can therefore be revealing in general, but the specific
numerical values obtained should be taken with some caution.

The purpose of the present paper is to draw attention to the fact that the
\HI\ produced from photodissociated \Htwo\ can also be a useful diagnostic for
determining physical conditions in PDRs.  In particular, the combination of
the 21~cm \HI\ and 2.6~mm CO(1--0) lines provides a means of independently
estimating $n$ and \Gzero\ in distant PDRs using ground-based radio astronomy
data.  These two radio lines form a diagnostic that is particularly useful in
the area of parameter space where \Gzero\ $\lesssim 100$, 10 \pcmcub $\lesssim
n \lesssim 10^5$ \pcmcub; this range of FUV flux and total density is
representative of most of the volume of the ISM in galaxies.

\section{PDR Models}
\label{models}

Our point of departure is the paper by \citet{kau99} that describes the
results of steady state computations on a specific, extensive set of PDR
models over a wide range of physical conditions.  These models are simple
one-dimensional semi-infinite slabs of gas with constant density $n = n(\HI) +
2n(\Htwo)$ of H nuclei subjected to an equivalent one dimensional flux of FUV
photons \Gzero\ measured in units of \pwr{1.6}{-3} ergs \pcmsq\ \ps\ over the
photon energy range 6--13.6 eV.\footnote{In these units, the average
interstellar radiation field over $4\pi$ sr is \Gzero\ $\approx 1.7$
(\citet{dra78}, see our Appendix \ref{app:ISRF} and Draine's Fig.\ 3).}  The
line brightness is calculated along a ray normal to the surface of the slab.
See \citet{kau99} for further details of their models.

\subsection{2.6~mm CO(1--0) Line Emission}

\begin{figure}[ht]
%\epsscale{0.8} % this figure is too big
\plottwo{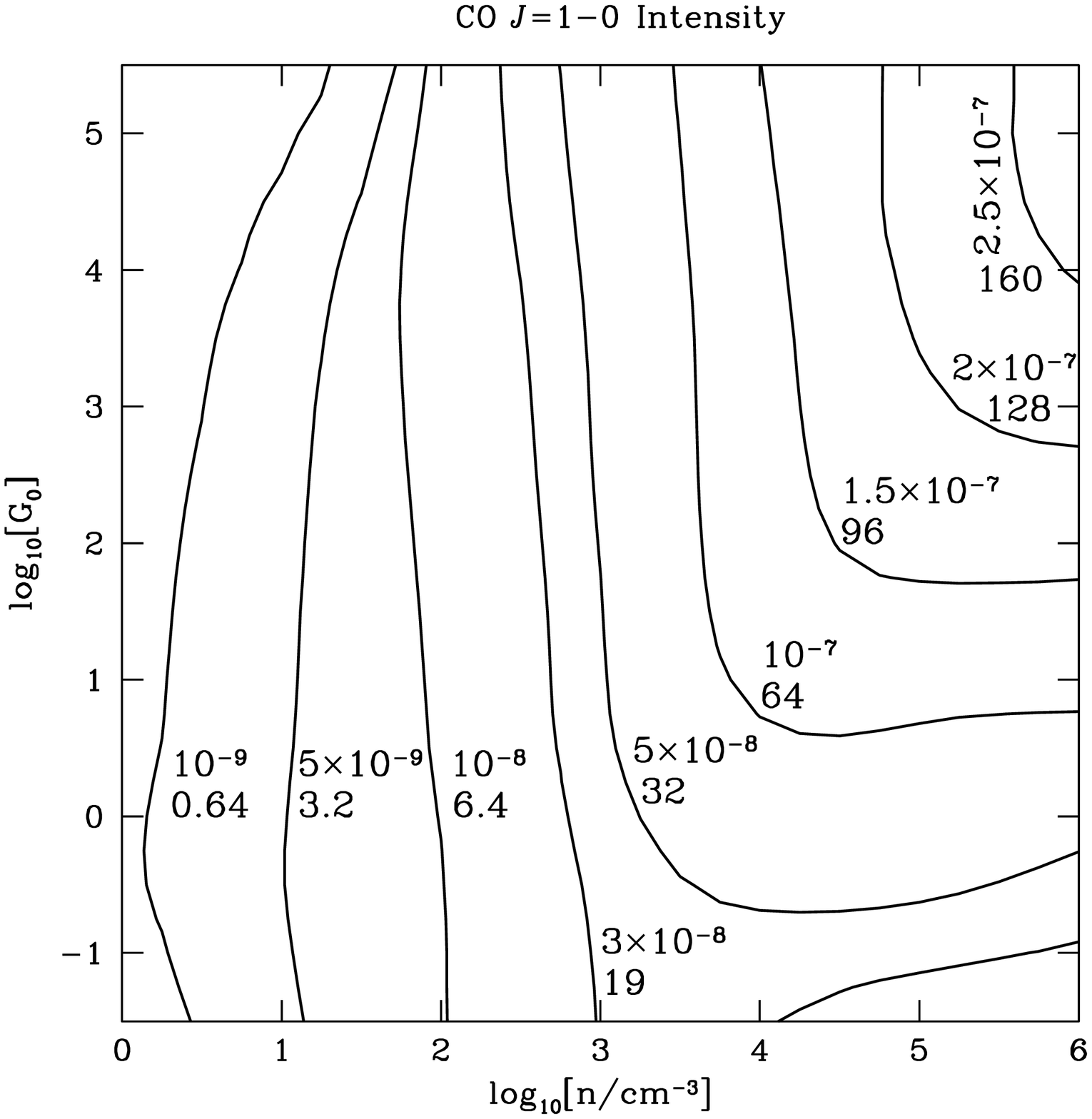}{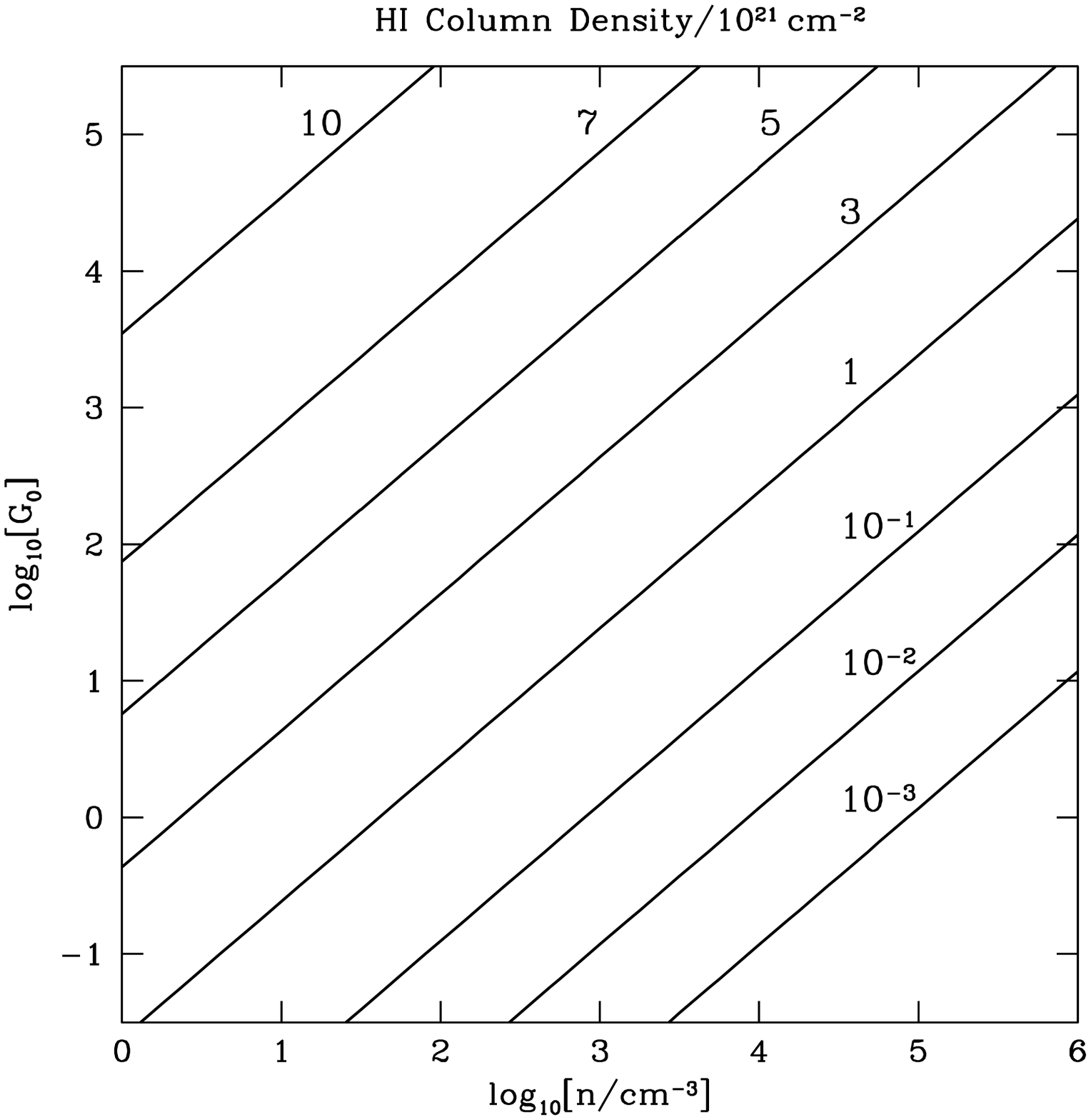}

\caption {a).  (left panel) CO(1--0) emission at 2.6~mm from the surface of
the model PDR as a function of the density $n$ and incident FUV flux \Gzero\
for the standard model parameters in \citet{kau99}.  The density $n = n_1 +
2n_2$ is the total density of H in atomic ($n_1$) and molecular ($n_2$) form
at the depth where the CO(1--0) line is formed.  At this point the gas is
essentially all \Htwo.  Contours of constant CO(1--0) emission $I_{\rm CO}$ are
shown, labeled in both ergs \pcmsq\ \ps\ \psr\ and in CO-observer units of K
\kmps.  The PDR is assumed to have the dust-to-gas ratio $\delta$ of the local
ISM near the Sun.  b).  (right panel) \HI\ column density $N(\HI)$ in units of
$10^{21}$ \pcmsq\ produced by photodissociation in the analytic approximation
of eq.  \ref{eqn:dissociate1}.  This approximation is preferred for
comparisons with observational data (see \S \ref{subsec:HI-line}).
\label{fig:modelCOHI}}

\end{figure}

The incident FUV flux dissociates and heats the cloud surface.  At some depth
into the cloud, the dissociation rate has decreased enough to allow CO
molecules to survive, and the density and heating rate are still high enough
to excite rotational line emission from these molecules.  Among many important
results, \citet{kau99} have computed the 2.6~mm CO(1--0) brightness, $I_{\rm
CO}$, emanating from the cloud surface; this is the most widely studied
molecular line in the ISM.  In Figure \ref{fig:modelCOHI}a we present their
results for $I_{\rm CO}$ as a contour diagram of constant surface brightness
(see their Fig.\ 11); we have extended the range of $n$ and \Gzero\ down to 1
and 0.03, respectively, and labeled the contours with their original
``theoretical'' units of ergs \pcmsq\ \ps\ \psr, as well as with ``CO
observer'' units of K \kmps.  This latter conversion is made according to:
\begin{equation} T_{\rm mb} \times \Delta V =
\frac{\lambda^3}{2 k_{\rm B}} I_{\rm CO} = 6.4 \times 10^8 \times I_{\rm CO} ,
\label{eqn:TmbCO}
\end{equation}
\noindent where $T_{\rm mb}$ is the measured emergent Rayleigh-Jeans
brightness temperature assuming the cloud is completely resolved,
$\Delta V$ is the profile velocity width in \kmps, $\lambda = 2.60$ mm, and
$k_{\rm B}$ is Boltzmann's constant.

We note that Figure \ref{fig:modelCOHI} assumes that the dust-to-gas ratio
$\delta$ is that of the local ISM near the Sun, and only one side of the slab
contributes to the observed emission.  A further caution concerns the value of
the density $n$ inferred from a specific observed value of the CO(1--0)
intensity in Figure \ref{fig:modelCOHI}a.  There is a general trend in most
PDR models to underestimate the CO abundance, so that the density inferred
from a given observed value of the CO(1--0) line is often higher than that
obtained from other independent diagnostics.  This is another example in which
comparisons among the different modeling efforts will be very useful.

The CO(1--0) surface brightness in Figure \ref{fig:modelCOHI}a is governed by
a complex interplay of processes involving the dissociation and reformation
of CO molecules, heating and cooling in the ISM, and the detailed excitation
and macroscopic radiative transfer of the CO(1--0) line.  By contrast, the
21~cm \HI\ line intensity is somewhat easier to compute, since in
general this line is optically thin and independent of the kinetic
temperature of the ISM; we can therefore ignore the physics of the 21~cm line
formation and simply compute the amount of \HI\ produced.

\subsection{21~cm \HI\ Line Emission}
\label{subsec:HI-line}

Incident FUV photons in the wavelength range 912--1108 \AA\ (13.6--11.2
eV) also dissociate the \Htwo\ on the surface of the GMC.\footnote{Photons
with energies as low as 6.6 eV can also contribute to the dissociation at
high FUV flux levels of \Gzero\ $\gtrsim 10^4$ \citep{shu78}.}  This process
was first described in the literature by \citet{ste67} and is by now well
known: it has been summarized by \citet{hol99}, who also provide several
other references.  The dissociated \Htwo\ creates a ``blanket'' of atomic gas
on the surface of the GMC; the \HI\ atoms recombine into \Htwo\ on dust grains
and are then returned to the gaseous ISM.  Approximate analytic expressions
for the steady state column density $N(\HI)$ of the atomic gas have been
derived by several authors under a variety of limiting assumptions (e.g.\
\citet{ste88, gol95}) and are discussed in more detail in Appendix
\ref{app:modelHIboth}.  Computations of $N(\HI)$ in the context of the
present ``standard model'' were first presented in \citet{wol90} (see their
Fig.\ 6).  An updated calculation using the latest code and the model
parameters in \citet{kau99} is presented as the dotted curves in Figure
\ref{fig:modelHIboth}.  The results differ slightly from those of
\citet{wol90}, owing to small differences in the assumed values of dust
abundance, atomic abundances, and a different \Htwo\ formation rate.  This
leads to changes in the depth of the \HI\ -- \Htwo\ transition, and hence to
different values of $N(\HI)$; the differences are, however, generally less
than about a factor of 2.

A problem arises in attempting to combine the computational results for
CO(1--0) with those for \HI.  The ``standard model'' used here is bathed in a
uniform flux of cosmic rays producing a constant ionization rate\footnote{This
parameter is unfortunately missing from Table 1 of \citet{kau99}.}  of
\pwr{1.8}{-17}\ \ps.  These cosmic rays also produce a faint extended
distribution of \HI\ atoms owing mainly to ion chemical reactions everywhere
in the cloud.  However, as we see below, such a faint extended distribution of
\HI\ will in general be removed from the observations by the data reduction
process, so we would prefer to use models for $N(\HI)$ \textit{without} a
cosmic-ray component.  Nevertheless, cosmic-ray ionization is important for
the chemistry in the model, thereby controlling to some extent the abundance
of CO molecules, so we ought to keep it in the computations of the CO(1--0)
line.

As a compromise we have chosen to use the approximate analytic expression for
$N(\HI)$ since it does not include any effects of cosmic rays.  This
approximation is presented in Appendix \ref{app:modelHIboth} where it is
compared with the standard computational model:  it is an excellent fit to the
model over the full range of parameter space important for this paper and
provides a better context in which to compare the theory with the
observations.  The adopted result for the \HI\ column is shown in Figure
\ref{fig:modelCOHI}b.

The conversion from \HI\ column density in Figure \ref{fig:modelCOHI}b
to 21~cm emission intensity in K \kmps\ is:
\begin{equation}
T_{\rm mb} \times \Delta V = 5.5 \times 10^{-19} \times N(\HI) ,
\label{eqn:TmbHI}
\end{equation}
\noindent for $N(\HI)$ in units of atoms \pcmsq, and assuming the optical
depth of the atomic gas to the 21~cm line radiation is small.

A noteworthy feature of Figure \ref{fig:modelCOHI}b is \textit{the constancy
of N(\HI) for a constant ratio of \Gzero/$n$} (see also eq.\
\ref{eqn:dissociate1}); the same \HI\ column density can be produced in high
FUV flux, high-density environments or in low FUV flux, low-density
environments.  In addition, \textit{the \HI\ column decreases with increasing
density at a given value of FUV flux}:  this occurs because the destruction
rate for \Htwo\ varies as $n$, but the formation rate varies as $n^2$ (since
the grain density is proportional to $n$).

\subsection{A Combined Model}

Figures \ref{fig:modelCOHI}a and \ref{fig:modelCOHI}b show that the contours
of constant CO(1--0) emission $I_{\rm CO}$ and constant \HI\ column density
$N(\HI)$ behave in a complementary fashion.  This suggests that over much of
the $n$--\Gzero\ plane an observational determination of $I_{\rm CO}$ and
$N(\HI)$ for the same PDR would permit independent estimates to be made of
both $n$ and \Gzero.  In Figure \ref{fig:modelboth} we show the two model
computations from Figures \ref{fig:modelCOHI}a and b superposed, along with
several additional contours and a ``box,'' which will be discussed further
below.

\subsubsection{Range of Validity of the Models}

The models become progressively less reliable in the top left corner of the
plots for both $I_{\rm CO}$ and $N(\HI)$.  This occurs because the assumption of a
steady state model becomes increasingly unreliable in the high $G_0$ -- low $n$
section of the diagram; in this region, radiation pressure causes grains to
drift with speeds of the same order as the gas turbulence speeds (see
\citet{kau99}, Fig.\ 2).  Furthermore, this is also the area of the diagram
where large \HI\ column densities are predicted; however, values of $N(\HI)$
in excess of a few $\times 10^{21}$ cm$^{-2}$ are not likely to be observed
since the 21~cm line may become optically thick in this regime:
\begin{eqnarray}
N(\HI)	& = & 1.82 \times 10^{18} \int_{- \infty}^\infty T_s \tau (v)
dv \nonumber \\
 & \rightarrow & 1.82 \times 10^{18} T_s \tau \Delta v \nonumber \\
 & \approx & {\rm\ a\ few} \times 10^{21} {\rm\ cm}^{-2}, \nonumber
\end{eqnarray}
\noindent for spin temperatures of $T_s \approx 50-100$ K, profile FWHMs
of $\Delta v \approx 4-8$ \kmps, and $\tau \approx 2-3$, values that
probably represent the extremes typical for the observations.

\subsubsection{Resolution Effects}

It should be emphasized that the present calculations of surface brightness
can be applied directly only to observations for which the sources of emission
are resolved.  In any realistic situation, the emission probably arises from
several PDR surfaces.  In this case, the density and FUV intensity derived
from our model using observations of $I_{\rm CO}$ and $N(\HI)$ represent
ensemble averages of the distribution of values for $n$ and \Gzero.
\citet{wol90} have shown that these averages are somewhat biased toward higher
FUV fields; bias toward high or low density depends on the cloud distribution.

\subsubsection{Interpretation of $n$}

The x~axis of Figure \ref{fig:modelboth} refers to the total volume density in
our computational model of an isopycnic cloud.  In a real GMC this is likely
to be close to the actual density at the depth in the cloud where the CO(1--0)
line is formed.  However, it is not likely to be representative of the density
of dissociated \HI.  This is because the various heating processes at the
surface of the GMC (including photodissociation) will lead to gradients in
temperature and, therefore, if the pressure is roughly constant, to gradients
in density.  For example, the model computations of \citet{all95} for
relatively low FUV fluxes\footnote{The symbol $\chi$ used by Allen et al.\ to
denote the FUV flux has very nearly the same value as we use here, $\chi =$
\Gzero/0.85 for irradiation over $2\pi$ sr.  See Appendix \ref{app:ISRF} for
a discussion of the various definitions of FUV values used in the literature
on PDRs.}  of \Gzero\ $\approx 0.1$ show that the density at the
CO(1--0)-emitting layer is about 5 times that at the \HI\ -- \Htwo\ interface.
Their Figure 4 shows that the CO emission emanates from a depth such that $A_V
\gtrsim 2$, where the temperature is just below 5 K (see their Fig.\ 3b),
whereas their Figure 2 shows the \HI\ -- \Htwo\ interface is effectively at
very small values of $A_V \approx 10^{-3}$, where the temperature is about 25
K.  We have calculated temperatures for higher FUV flux levels \Gzero\ = 3:
the surface temperatures are higher in this case, $\approx 60$ K at the \HI\
-- \Htwo\ interface, and $\approx 7.5$ K at the CO emission layer, for a ratio
of $\approx 8$.  This ratio will continue to rise for yet higher photon
fluxes, since the increased surface heating will result in further increases
in surface temperature, but the increase in temperature deeper in the clouds
where the CO(1--0) emission is formed will be less rapid because of more
effective cooling.  If the density and temperature are linked by an isobaric
condition, the density observed in the bulk of the \HI\ may be a factor of
5--10 lower than that inferred for the CO(1--0) line.

It is clear from this discussion that constant-density models for PDRs such
as the one we have assumed here are not likely to be very realistic.  More
sophisticated models, for example assuming pressure equilibrium, or a full
computation with various forms of micro- and macro-turbulent pressure
(e.g., \citet{wol93}) over the range of the \Gzero--$n$ parameter space
may provide more realistic results and ought to be explored.

\section{Discussion}

\subsection{Complementarity}

Figure \ref{fig:modelboth} clearly shows the complementarity of 21~cm \HI\ and
2.6~mm CO(1--0) radio line emission as diagnostics of PDRs.  Over the range of
validity of our calculations, the CO(1--0) line brightness $I_{\rm CO}$ depends
mostly on the volume density of the gas for $n \lesssim 10^3$ \pcmcub, no
matter how intense the FUV flux.\footnote{The insensitivity of $I_{\rm CO}$ to
the FUV flux comes about because increasing \Gzero\ merely leads to more
photodissociation of the surface CO, and the CO-emitting layer retreats deeper
into the static cloud where the temperatures are about the same as they were
for the case of lower \Gzero.}  On the other hand, for constant $n$ the \HI\
brightness increases nearly linearly with increasing FUV fluxes up to \Gzero\
$\approx 10-100$, becoming logarithmic for higher FUV fluxes.  Accordingly,
the combination of measurements of $N(\HI)$ and $I_{\rm CO}$ from the same PDR
can provide unique values for $n$ and \Gzero\ over a substantial range in the
diagram.

\subsection{Identifying the Relevant \HI\ and CO Emission}

How can extraneous emission from gas that lies far outside the cloud be
identified?  The spatially layered structure of PDRs provides a direct clue,
as has been convincingly shown e.g.\ by the observations of the Orion Bar
region (see Fig.\ 2 in \citet{hol99}) on the $\sim 1$ pc scale.  A more or
less edge-on viewing angle is required for a certain identification, and one
ought to remove any extended emission that is associated with other clouds
along the line of sight or produced by other excitation mechanisms.  Galaxies
that are viewed at low to intermediate inclination angles are favorably
oriented for this identification to work, and \cite{all97} and \cite{smi00}
have shown that the signature morphology can be found at 100 pc scales in the
highest resolution \HI\ images of nearby galaxies M81 and M101.  As to the CO,
it is generally assumed that this arises in PDRs, although the morphological
signature is not always clear.

\subsection{G216~-2.5: An Example}

Figure \ref{fig:modelboth} provides a method for determining $n$ and \Gzero\
for a specific PDR if $N(\HI)$ and $I_{\rm CO}$ can be determined
observationally.  As an example, we consider the Galactic object G216~-2.5
(also called ``Maddalena's Cloud'') which has been proposed to be a large PDR
in the Galaxy by \citet{wil96}.

\subsubsection{\HI\ Column Density}

Determining the column density of that part of the \HI\ associated with
G216~-2.5 is made quite difficult by confusion from unassociated \HI\
superposed along the line of sight through the Galaxy.  Several estimates can
be made from the single-dish observations reported by \citet{wil96}.  First,
and perhaps simplest, we estimate that fraction of the total averaged \HI\
profile shown in their Figure 1 that is in the velocity range of the CO(1--0)
emission (also shown on the same figure); the \HI\ profile shows three
overlapping peaks and integrates to $\approx 48$ K $\times \; 58$ \kmps $\approx
2800$ K \kmps = \pwr{5.1}{21} \pcmsq.  The averaged $I_{\rm CO}$ profile
corresponds to the central \HI\ peak and is $8.5$ \kmps\ wide \citep{mad85},
so the average amount of \HI\ column associated with the CO emission is, by
this estimate, probably not more than $\approx (8.5/58) \times$ \pwr{5.1}{21}
= \pwr{7.5}{20} \pcmsq.  In their \S 3.2.2, \citet{wil96} carry out a more
involved analysis with area integrations of the \HI\ in the velocity range
16--38 \kmps\ and conclude that the ``excess'' \HI\ associated with the PDR is
$\approx$ \pwr{2}{20} \pcmsq; the uncertainty in this method is approximately
a factor of 2 (R.\ Maddalena 2002, private communication).  A value in the
range \pwr{1-8}{20} \pcmsq\ may therefore be considered typical for this
cloud, with some preference for the lower end of that range.

\subsubsection{CO(1-0) Intensity}

From Figure 1 of \citet{wil96}, the total $I_{\rm CO}$ profile over the whole
cloud integrates to $\approx 7.1$ K \kmps.  The CO(1--0) map for the cloud
integrated over the same velocity range of 16--38 \kmps\ as the \HI\
is shown in Figures 2 and 4 of \citet{wil96}, based on earlier data of
\citet{mad85}.\footnote{Note that the $l - b$ maps in \citet{wil96} are
similar in shape but brighter by about a factor of 2 when compared to the
earlier map of $I_{\rm CO}$ in Fig.\ 2 of \citet{mad85}.  This is not likely
to be an effect of differences in the velocity range over which the data
have been integrated, since that range is smaller than the range of 15--40
\kmps used by Maddalena \& Thaddeus.  We use the more recent results of
Williams \& Maddalena.} From Figure 2 of \citet{wil96} the CO(1--0)
brightness ranges from 3--7 contours with a few positions reaching 10
contours.  A range of 3--8 contours or 6--16 K \kmps\ therefore seems
representative of the CO data, again with some preference for the lower end
of that range.

\subsubsection{$n$ and \Gzero\ for G216~-2.5}

The range of values for $I_{\rm CO}$ and $N(\HI)$ obtained above is plotted as a
black box on our Figure \ref{fig:modelboth}.  In the context of our
model, these values describe gas with a density range of \pwr{0.7-5}{2}
\pcmcub\ and an incident FUV flux of 0.1--7 of the standard value \Gzero\
near the Sun, with likely values being approximately \pwr{2}{2}
\pcmcub\ and 0.8, respectively.

Although no independent measurements of the total volume density are available
for G216~-2.5, the value we have obtained is in the range generally indicated
for GMCs in the Galaxy ($n = n(\HI) + 2n(\Htwo) \approx 2n(\Htwo) \sim 2
\times 50$ \pcmcub, e.g., \citet{bli93}).  As to the FUV flux, \citet{wil96}
estimated \Gzero\ $\approx 1$ from what is known about the two nearby young
stars identified in their study as likely to be responsible for the
photodissociation.  Our model thus agrees well with the observations.

\subsection{Approximate Detection Limits for \HI\ and CO(1--0)}
\label{subsec:DetLims}

\begin{figure}[ht]
%\epsscale{0.8} % this figure is too big
\plotone{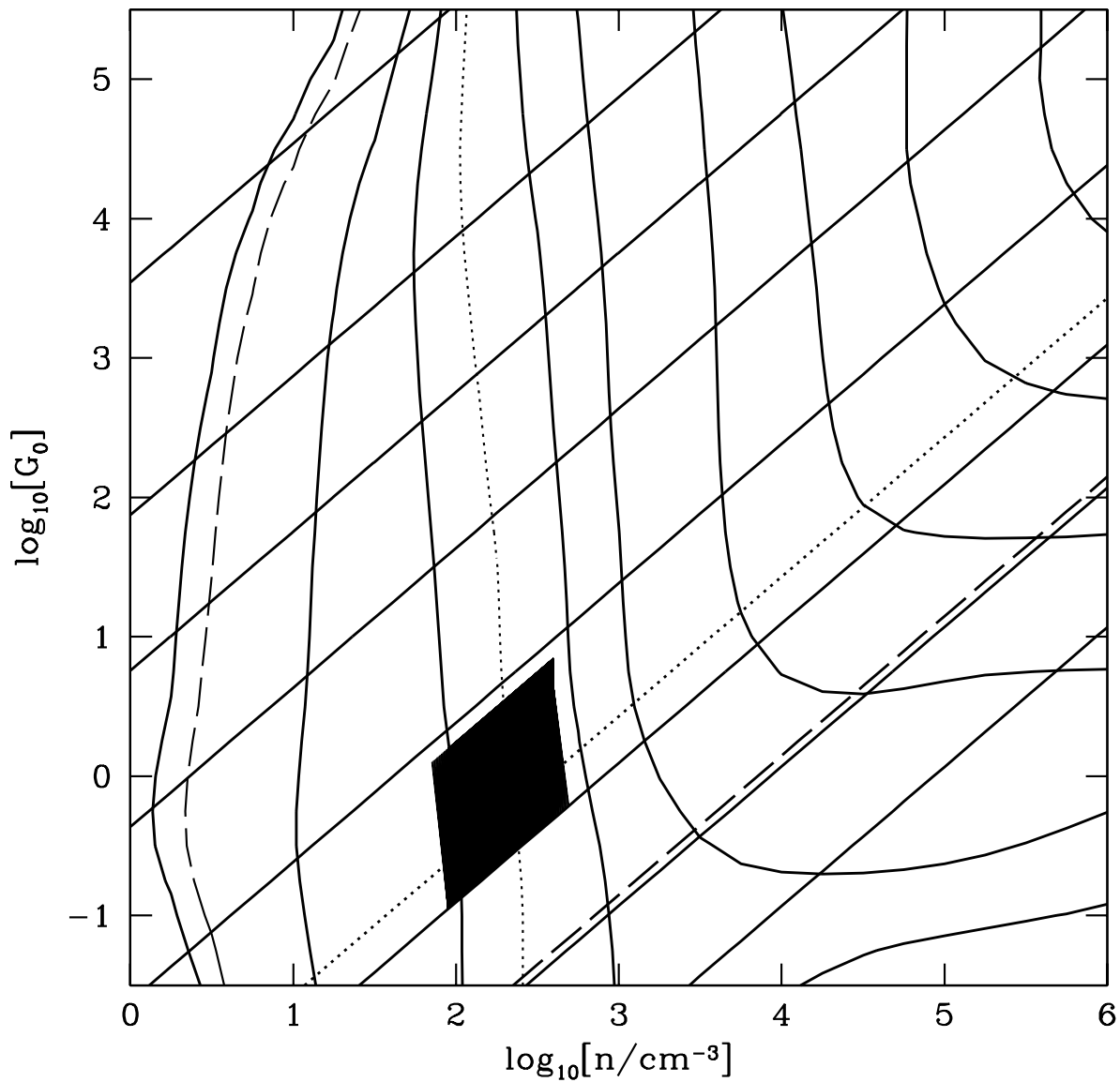}

\caption {Combined plot from the model computations of CO(1--0) emission
$I_{\rm CO}$ in Fig.\ \ref{fig:modelCOHI}a and $N(\HI)$ in Fig.\
\ref{fig:modelCOHI}b.  The contour labels, identical with Fig.\
\ref{fig:modelCOHI}, are omitted here for clarity.  Shown are approximate
observational limits for single dish (\textit{dashed lines}) and
interferometer imaging synthesis data (\textit{dotted lines}).  These limits
are described in the text \S \ref{subsec:DetLims}.  The black ``box'' shows
the approximate range allowed by the observations of G216~-2.5.
\label{fig:modelboth}}

\end{figure}

\begin{figure}[ht]
\epsscale{0.8} % this figure is too big
\plotone{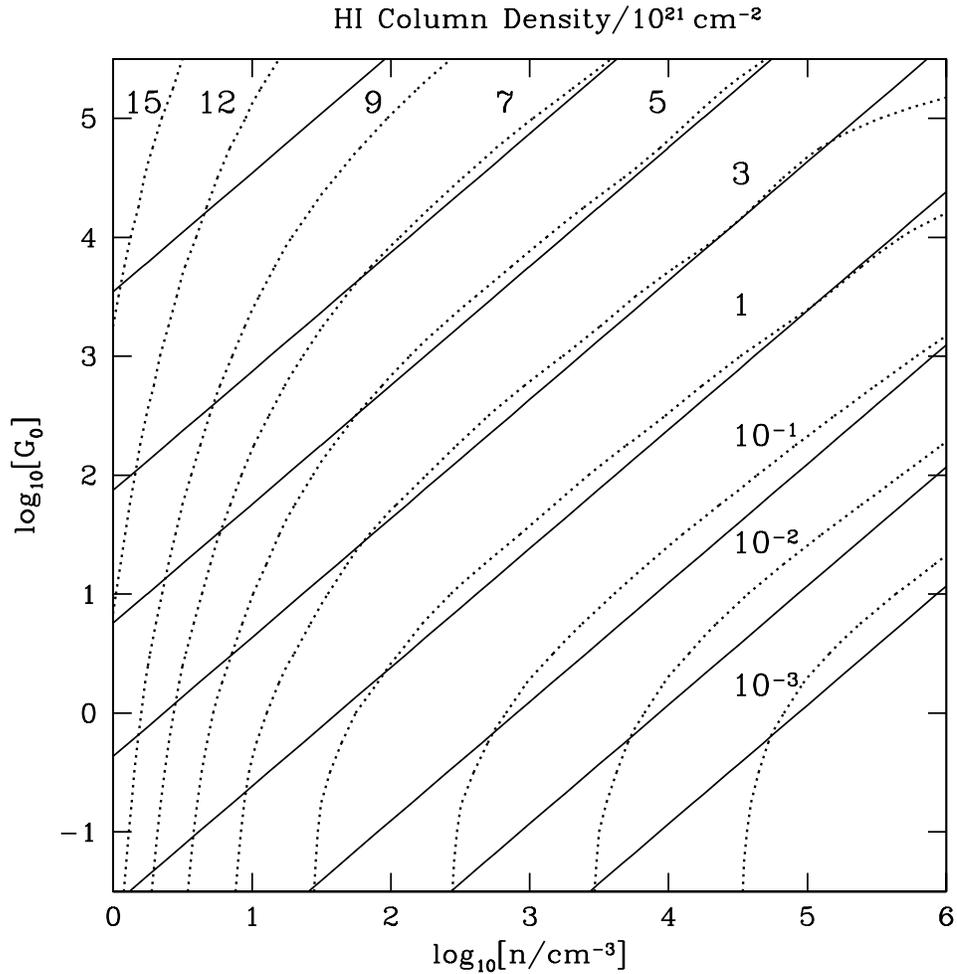}

\caption {Contours of constant \HI\ column density $N(\HI)$ in units of
$10^{21}$ \pcmsq\ in the PDR as a function of the density $n$ and incident FUV
flux \Gzero\ for the standard model parameters in \citet{kau99}
(\textit{dotted lines}), and for the analytic approximation of eq.\
\ref{eqn:dissociate1} (\textit{solid lines}).  The labeled contour values are
for the numerical model; the contours for the analytic model are as in Fig.\
\ref{fig:modelCOHI}.  \label{fig:modelHIboth}}

\end{figure}

The practical detection limits for the \HI\ 21 cm line are $\approx 5$ and
$\approx 100$ K \kmps\ for single-dish survey and interferometer array imaging
observations, respectively, corresponding to $N(\HI) \approx$ \pwr{1}{19} and
\pwr{2}{20} \pcmsq.  These are shown as the dashed (single-dish survey) and
dotted (interferometer array) lines for $N(\HI)$ on Figure
\ref{fig:modelboth}.  The corresponding values for the 2.6~mm CO line are
$\approx 1$ K \kmps\ (dashed line:  single dish survey) and $\approx 10$ K
\kmps\ (dotted line:  interferometer array).  In all cases, this assumes that
the gas clouds are resolved; otherwise these limits need to be increased
further by the ratio of the beam area to the cloud area.

Locating the detection limits on our Figure \ref{fig:modelboth}, we can
conclude that the range in parameter space over which the combination of \HI\
and CO(1--0) can provide useful constraints on $n$ and \Gzero\ is given
approximately by \Gzero\ $\lesssim 100$, 10 \pcmcub\ $\lesssim n \lesssim
10^5$ \pcmcub.  These boundaries are set by the decreasing sensitivity of the
\HI\ emission at high FUV flux levels ($N(\HI)$ increases only logarithmically
for high \Gzero\ at constant $n$) and by observational detection limits
($I_{\rm CO}$ disappears at low $n$ owing to insufficient excitation of the
transition; \HI\ disappears at high $n$ as the gas stays mostly molecular).

\subsubsection{Working with Synthesis Imaging Data}

Many observers who report synthesis imaging (interferometer array) data in
the literature use units of Jy beam$^{-1}$ instead of K.  The required
conversion factors are (unfortunately) telescope dependent; Appendix
\ref{app:conversions} provides an approximate recipe for these conversions.

\subsection{Conclusions}

We have shown that a combination of observations in the 21~cm line of \HI\ and
the 2.6~mm line of CO can provide a useful diagnostic for physical conditions
in distant PDRs.  Both of these lines are readily observed in the Galaxy and
in nearby galaxies using ground-based single-dish and interferometer-array
radio telescopes.  The useful range in parameter space for this combined radio
line diagnostic is approximately bounded by \Gzero\ $\lesssim 100$, 10 \pcmcub
$\lesssim n \lesssim 10^5$ \pcmcub, a range that covers a large fraction of
the volume of the ISM in galaxies.  The unique layered morphology of the
emission in PDRs provides a means of identifying that part of the \HI\ and CO
emission that is related when that morphology is observable, but confusion
along the line of sight makes this separation difficult in the Galaxy.  This
method will be especially useful for the interpretation of high-resolution
\HI\ and CO synthesis imaging data in nearby galaxies where the PDR morphology
can be more easily identified.\footnote{However, in this ``extragalactic''
case, the effects of beam-smearing are severe and must be carefully considered
in relating the observational parameters of the models to the measured
brightnesses and profile velocity widths.}  Improvements in the model
parameters and in the details of the computations are anticipated in the near
future that will permit even more precise values of \Gzero\ and $n$ to be
obtained.

\acknowledgements

We are grateful to Bruce Draine and David Hollenbach for discussions on the
physics of PDRs.  David also provided very helpful comments on earlier
drafts of this paper.  The referee offered many useful suggestions that have
also been incorporated.

\appendix	% this apparently reverts to one column...

\clearpage

\section{COMPARISON OF ANALYTIC AND NUMERICAL MODELS FOR \HI\ PRODUCTION
IN PDRs} \label{app:modelHIboth}

The \HI\ column density in a PDR is calculated with the same physics used to
determine the excitation of the \Htwo\ near-infrared fluorescence lines.  A
logarithmic form for the analytic solution to the equation for
formation--destruction equilibrium was first given by \citet{ste88} (see also
eq. A2 in \citet{hol91}\footnote{David Hollenbach has kindly pointed out
to us that there is a typographical error in eq.\ A1 of their paper; the
right side should read $I_0 (\beta/N_2^{0.5}) n_2 e^{-KN}$.}  and eq. 3
in \citet{gol95}).  The model is a simple semi-infinite slab geometry in
statistical equilibrium with FUV radiation incident on one side.  The solution
gives the steady state \HI\ column density along a line of sight perpendicular
to the face of the slab as a function of $\chi$, the incident UV intensity
scaling factor (see Appendix \ref{app:ISRF}), and the total volume density $n$
of H nuclei.  Sternberg's result is:

\begin{equation}
N(\HI)={1 \over \sigma} \times \ln{\left[ {{D G} \over {R n}}\chi
+ 1 \right]},
\label{eqn:nhi}
\end{equation}

\noindent where;

\begin{tabular}{rcp{2.5in}rcp{2.0in}}
$D$ & = & the unattenuated \Htwo\ photodissociation rate in the average ISRF,
& $R$ & = & the \Htwo\ formation rate coefficient on grain surfaces, \\
$\sigma$ & = & the effective grain absorption cross section
per H nucleus in the FUV continuum,
& $\chi$ & = & the incident UV intensity scaling factor, \\
$N(\HI)$ & = & the \HI\ column density,
& $n$ & = & the volume density of H nuclei. \\
 & & & & & \\
\end{tabular}

Equation \ref{eqn:nhi} has been developed using a simplified three-level model
for the excitation of the \Htwo\ molecule and is applicable for low-density
($n \lesssim 10^4$ \pcmcub), cold (T $\lesssim 500$ K), isothermal, and static
conditions, and neglects contributions to $N(\HI)$ from ion chemistry and
direct dissociation by cosmic rays.  The quantity $G$ here (not to be confused
with \Gzero\ used previously) is a dimensionless function of the effective
grain absorption cross section $\sigma$, the absorption self--shielding
function $f$, and the column density of molecular hydrogen $N_2$:
\begin{displaymath}
G = \int_0^{N_2} \sigma f e^{-2\sigma N_2^\prime} {\rm d}N_2^\prime.
\end{displaymath}
The function $G$ becomes constant for large values of $N_2$ due to
self--shielding \citep{ste88}.  Using the parameter values in this equation
adopted by \citet{mad93}, we have:
\begin{displaymath}
N(\HI) = \pwr{5}{20} \times \ln [1+ (90\chi/n)]
\end{displaymath}
\noindent where $n$ is in \pcmcub.  This is a steady state model, with \Htwo\
continually forming from \HI\ on dust grain surfaces, and \HI\ continually
forming from \Htwo\ by photodissociation.  In order to compare this result
with the ``standard model'' computations we need to relate $\chi$ to the
\Gzero\ used by \citet{kau99}; this is because \citet{ste88} and \citet{kau99}
use different normalisations for the FUV flux (see Appendix \ref{app:ISRF}).
When distributed sources illuminate an FUV-opaque PDR over $2\pi$ sr, the
conversion is $\chi = G_0/0.85$ (see Footnote 7 in \citet{hol99}), resulting
in $90 \chi /n = 106 G_0/n$.

We have fitted the analytic expression for $N(\HI)$ to the model computations in
the range in which cosmic ray dissociation is not a major contributor, roughly for
\Gzero\ $\gtrsim 1$, $n \gtrsim 10$ \pcmcub.  The result is that no consistent
improvement is obtained by using any value for the coefficient of \Gzero\ other
than the value 106 deduced above, although a modest improvement is obtained by
using a slightly larger value for the leading coefficient in the equation,
\pwr{7.8}{20}, corresponding to a value of \pwr{1.3}{-21} cm$^2$ for the
effective grain absorption cross section.  With these small adjustments, our
final equation is:

\begin{equation}
N(\HI) = \pwr{7.8}{20} \times \ln [1+ (106G_0 /n)] {\rm ~cm}^{-2},
\label{eqn:dissociate1}
\end{equation}

\noindent resulting in an $r^2$ value for the fit that is everywhere in excess
of 99\% over the fitted range of $n$ and \Gzero.

In Figure \ref{fig:modelHIboth} we show values from equation
\ref{eqn:dissociate1} plotted as solid lines together with dotted contour
lines from our standard numerical model.  The agreement is generally good over
much of the $n$--\Gzero\ parameter space of interest here; differences occur
mainly in the top left corner of the diagram, and at low values of FUV flux.
In the top left corner the analytic formula under-predicts the amount of \HI\
column density computed from the standard model by about 30\% owing to \HI\
production by ion chemistry reactions such as $H_2^{+} + H_2 \rightarrow
H_3^{+} + H$, $HCO^{+} + e^{-} \rightarrow CO + H$, and $PAH^{-} + H^{+}
\rightarrow PAH + H$ (where PAH is a polycyclic aromatic hydrocarbon), which
are important at high \Gzero\ and low $n$.  At values of \Gzero\ $\lesssim 1$
the contours of $N(\HI)$ for the numerical model become vertical; this is
because the standard model includes a low level of cosmic ray ionization (see
main text) which contributes a small amount of \HI\ even for \Gzero\ = 0.

%\clearpage

\section{DEFINITIONS OF THE MEAN INTERSTELLAR RADIATION FIELD}
\label{app:ISRF}
		
\cite{kau99} characterized the emission from photodissociation regions (PDRs)
in terms of the ``strength'' of the FUV radiation field, \Gzero, illuminating
the medium at energies between 6 and 13.6 eV.  Numerical values for \Gzero\
were presented in units of the ``Habing Field,'' following earlier PDR
modeling by \cite{tie85}, who expressed \Gzero\ in units of the ``equivalent
\cite{hab68} flux of \pwr{1.6}{-3} ergs cm$^{-2}$ s$^{-1}$ appropriate to the
average interstellar medium.'' However, other characterizations of the FUV
background also appear in the literature for the ``typical'' and local ISM.
\cite{dra78} has developed an analytic fit to many of these, which he defined
as the ``standard UV background'' from 5--13.6 eV.

Although numerical values for the \cite{hab68} and \cite{dra78}
representations of the average interstellar background can be interconverted,
Draine's Figure 3 makes it evident that they have different spectral shapes.
Furthermore, the ``strength'' of the interstellar FUV field is described in
the literature by several different quantities (e.g., energy density, photon
intensity, energy flux) using a variety of units.  Some authors use a single
symbol to report normalized values of their field descriptor, while others
specifically represent the normalized quantity by a ratio.  Accordingly, the
purpose of this appendix is to summarize the approach taken by Habing and
Draine in developing their descriptions of the ISRF at ultraviolet
wavelengths and to interrelate the representations of that field adopted by
several authors whose work is germane to our paper.

\cite{hab68} calculated radiation energy density values, $u_{\lambda}$, at
1000, 1400, and 2200 \AA\ which were averaged ``over a considerable area'' in
space, using bandwidths of 912--1040, 1300--1450, and 1900--2400 \AA,
respectively.  His recommended values for the averaged energy density $\langle
u_{\lambda} \rangle$ were

\begin{eqnarray}
\langle u_{\lambda} \rangle & = & 40 \times 10^{-18}
\mathrm{\ ergs\ cm^{-3}\ \AA^{-1}\ at\ 1000\ \AA\ (11.9-13.6\ eV)}
\nonumber \\
 & = & 50 \times 10^{-18}
\mathrm{\ ergs\ cm^{-3}\ \AA^{-1}\ at\ 1400\ \AA\ (8.6-9.5\ eV)} \\
 & = & 30 \times 10^{-18}
\mathrm{\ ergs\ cm^{-3}\ \AA^{-1}\ at\ 2200\ \AA\ (5.2-6.5\ eV)} \nonumber
\label{eqn:ulambda}
\end{eqnarray}

\noindent However, Habing's results depend strongly on assumptions made
concerning the wavelength dependence of interstellar extinction and the presence
of diffuse galactic light.  \cite{wit73} noted that these are ``two areas for
which there existed little or no observational data at the time of his work.''

\cite{dra78} converted these and other existing calculations of the interstellar
UV energy density into a quantity that he defined as the angle-averaged photon
flux

\begin{equation}
F(E) \equiv \lambda^3 u_\lambda / (4 \pi h^2 c)
\mathrm{\ photons\ cm^{-2}\ s^{-1}\ sr^{-1}\ eV^{-1}}
\label{eqn:photflux}
\end{equation}

\noindent where $E$ is the energy, $h$ represents Planck's constant, and $c$ is
the speed of light.  The resulting values and a few observational data points
are plotted on his Figure 3, together with an analytical expression that is ``in
good agreement with all of the above results over the range 5--13.6 eV.'' That
expression,

\begin{equation}
F_D(E) = 1.658 \times 10^6 E - 2.152 \times 10^5 E^2 + 6.919 \times 10^3 E^3
\mathrm{\ photons\ cm^{-2}\ s^{-1}\ sr^{-1}\ eV^{-1}}
\label{eqn:fitphotflux}
\end{equation}

\noindent where $E$ is measured in eV, was referred to by Draine as
the ``standard UV background'' in the ISM.  Since it is computed on a
``per steradian'' basis, this quantity is actually an angle-averaged photon
intensity.

We have used equation \ref{eqn:photflux} to convert the \cite{hab68} energy
density values to angle-averaged photon intensities (denoted $F_H(E)$) at the
three wavelengths reported in his paper, and they agree well with the
corresponding points on Figure 3 of \cite{dra78}.  We have also computed
values for Draine's ``standard'' UV background at the same wavelengths, using
equation \ref{eqn:fitphotflux}.  At 1000 \AA, $F_H(E) =$ \pwr{3.875}{5}
$\mathrm{\ photons\ cm^{-2}\ s^{-1}\ sr^{-1}\ eV^{-1}}$, $F_D(E) = $
\pwr{6.627}{5} (in the same units), and $F_H(E) / F_D(E) = 0.585$.  Thus, the
photon intensity computed from Draine's fit at this wavelength is 1.710 times
that found by converting Habing's calculated energy density to these units.
This result is identical to the value computed by \cite{dra96} at 1000 \AA,
and (to one decimal point) to the value given in footnote 7 of \cite{hol99}.
The same result could also be obtained by first converting Draine's photon
intensity to an energy density for direct comparison to Habing's (1968)
calculation.  \cite{dra96} chose Habing's result at 1000 \AA\ as the basis
for normalization because in ``neutral regions, pumping of \Htwo\ is primarily
effected by far-ultraviolet photons in the 1110--912 \AA\ range''.  The energy
flux over $4 \pi$ sr at 1000 \AA\ is $\mathsf{F} = 4 \pi E F_H(E) =$
\pwr{1.2}{-3} ergs cm$^{-2}$ s$^{-1} = 4 \pi \bar{I}$, where $\bar{I}$ is the
mean intensity.

The values for these quantities at 1400 \AA\ are $F_H(E) =$ \pwr{1.329}{6},
$F_D(E) =$ \pwr{2.611}{6}, and $F_H(E) / F_D(E) = 0.509$.  In this case,
Draine's fit to the angle-averaged photon intensity is 1.965 times the
converted Habing result, and the energy flux is $\mathsf{F} = 4 \pi E F_H(E)
=$ \pwr{2.1}{-3}.  At 2200 \AA, $F_H(E) =$ \pwr{3.094}{6}, $F_D(E) =$
\pwr{3.748}{6}, and $F_H(E) / F_D(E) = 0.826$.  The ``standard'' Draine result
is 1.211 times that calculated by Habing at this wavelength, and the energy
flux $\mathsf{F} = 4 \pi E F_H(E) =$ \pwr{2.0}{-3} .

Draine's standard curve can be integrated from 6 to 13.6 eV to find the total
photon intensity, i.e., \pwr{1.547}{7} photons cm$^{-2}$ s$^{-1}$ sr$^{-1}$.
When this result is multiplied by the average energy per photon (i.e.,
\pwr{1.407}{-11} ergs) over that interval, the Draine flux over $4 \pi$ sr is
$\mathsf{F}_D =$ \pwr{2.7}{-3} ergs cm$^{-2}$ s$^{-1}$.  If we assume that the
photon intensity ratio found earlier in the 11.9--13.6 eV band, e.g., $F_H(E)
/ F_D(E) = 0.585$, applies over the entire FUV spectrum (i.e., 6--13.6 eV),
then $\mathsf{F}_H =$ \pwr{1.6}{-3} ergs cm$^{-2}$ s$^{-1}$.  This is the
equivalent Habing flux.

Since both ultraviolet normalization procedures are still used in the
literature, confusion can arise when comparing descriptors of the ISRF.
\cite{kau99} represent that field by an ``incident FUV flux'', \Gzero, that is
already normalized by the equivalent Habing flux of the average ISM (see their
Table 1).  Normalization to the Habing field was also used by \cite{hol99} and
\cite{wol90}.  The use of this normalization in these models is purely
historical, having been incorporated into early versions of the underlying
code in support of PDR modeling begun in the 1980's (M.\ Wolfire 1999,
private communication).

In contrast, \cite{ste88} and \cite{ste89, ste95} compute the ``UV 
photon intensity'' (i.e., $\chi I_{\nu}$) by introducing a ``UV intensity
scaling factor'' $\chi$ which is defined 
to be unity for Draine's fit to the average interstellar field intensity
$I_{\nu}$.  In \cite{ste88} and \cite{ste89}, the \cite{dra78}
standard result is re-expressed as

\begin{equation}
I_{\nu} = \frac{1.068 \times 10^{-4}}{\lambda}
	- \frac{1.719 \times 10^{-2}}{\lambda^2}
	+ \frac{6.853 \times 10^{-1}}{\lambda^3}
\mathrm{\ photons\ cm^{-2}\ s^{-1}\ Hz^{-1}},
\label{eqn:drafreq}
\end{equation}

\noindent where $\lambda$ is expressed in nanometers.  When Sternberg's
representation of Draine's standard background (i.e., eq.
\ref{eqn:drafreq}) is integrated from 6 to 13.6 eV, and the resulting value,
\pwr{1.942}{8} photons cm$^{-2}$ s$^{-1}$, is multiplied by the average
energy per photon, the total FUV flux over $4 \pi$ sr is $\mathsf{F_S} = $
\pwr{2.7}{-3} ergs cm$^{-2}$ s$^{-1}$.  Once again, we find that $\mathsf{F}_H
=$ \pwr{1.6}{-3} provided the ratio $F_H(E) / F_D(E) = 0.585$ found earlier
between 11.9 and 13.6 eV (i.e., 912--1040 \AA) is assumed to apply over the
entire FUV band.

Equation \ref{eqn:drafreq} can be obtained from equation \ref{eqn:fitphotflux}
by multiplying the latter by $4\pi$ sr and \pwr{4.136}{-15} eV Hz$^{-1}$,
provided $E$ is replaced by $hc/\lambda = 1.240 \times 10^{-3}/\lambda_{nm}$
eV.  Thus,

\begin{equation}
I_{\nu} = 1.654\pi \times 10^{-14} F_D(hc/\lambda_{nm}).
\label{eqn:drafreq2}
\end{equation}

We note that equation \ref{eqn:drafreq} represents a photon flux rather than
the photon intensity, since it has been averaged over solid angle.
\cite{ste95} corrected this problem in their equation A1 by multiplying
equation \ref{eqn:drafreq} by $1/4\pi$, resulting in corrected units of
photons cm$^{-2}$ s$^{-1}$ Hz$^{-1}$ sr$^{-1}$.  When this correction is made,
equation \ref{eqn:drafreq2} becomes $I_{\nu} = 4.136 \times 10^{-15}
F_D(hc/\lambda_{nm})$.

\cite{mad93} explicitly introduced field normalization through the ratio
$\chi_{UV}/\chi_0$, where $\chi_{UV}$ is defined as the UV field intensity,
although they also called it an FUV energy flux.  Unfortunately, the subscript
``UV'' is dropped later in their paper where the ratio is shown as
$\chi/\chi_0$, increasing the potential for equating the numerator
(incorrectly) with Sternberg's parameter $\chi$ upon casual inspection.  The
divisor, $\chi_0 =$ \pwr{2.0}{-4} ergs cm$^{-2}$ s$^{-1}$ sr$^{-1}$, was
determined from the ``UV field'' results in the solar neighborhood presented
by \cite{dra78}.  Although not explicitly stated when $\chi_0$ was defined,
their introduction identifies the band of interest as extending from 912 -
2000 \AA (i.e., 6.2 - 13.6 eV).  Over $4\pi$ sr, the flux corresponding to the
Madden et al.\ value of $\chi_0$ is $\mathsf{F}_M =$ \pwr{2.5}{-3} ergs
cm$^{-2}$ s$^{-1}$, and $\mathsf{F}_H =$ \pwr{1.5}{-3} if we continue to
assume that $F_H(E) / F_D(E) = 0.585$.

The key issue for this paper is how to relate the dimensionless field
descriptor chosen by \cite{kau99}, i.e., \Gzero, to the dimensionless quantity
used by \cite{ste88}, $\chi$, or by \cite{mad93}, $\chi_{UV}/\chi_0$.  Since
$\chi$ was normalized by Draine's standard background between 6 and 13.6 eV,
and $\chi_0$ was computed by integrating the same curve over a similar energy
range (the lower limits may differ by 0.2 eV), $\chi_{UV}/\chi_0$ is
equivalent to $\chi$.  We illustrate this circumstance by comparing the flux
computed earlier from Sternberg's representation of Draine's standard curve
(i.e., from eq.\ \ref{eqn:drafreq}), given by $\mathsf{F}_S =$ \pwr{2.7}{-3}
ergs cm$^{-2}$ s$^{-1}$, with the flux computed (using fewer significant
figures) from the value of $\chi_0$ used by \cite{mad93}.  The latter is
$\mathsf{F}_M =$ \pwr{2.5}{-3} ergs cm$^{-2}$ s$^{-1}$.  This conclusion is
not affected by whether or not the \cite{ste88} expression is multiplied by a
solid angle factor (i.e., $1/4\pi$) provided his numerator is multiplied by
the same factor.

The dimensionless parameter \Gzero\ $= [4\pi \int F(E) dE] / [4\pi \int F_H(E)
dE]$ is a flux ratio that is computed over the same spectral band as $\chi_0$,
where the denominator $\int F_H(E) dE = 0.585 \int F_D(E) dE$ in the highest
energy band (i.e., 11.9--13.6 eV) addressed by \cite{hab68}.  Accordingly,
over that band \Gzero\ $= 1.710 \int F(E) dE / \int F_D(E) dE$ .  Since
$\chi_0$ (with units of ergs cm$^{-2}$ s$^{-1}$ sr$^{-1}$) represents an
intensity that is referenced to Draine's standard curve, it can be written as
$\int F_D(E) dE$, whereby \Gzero $= 1.710 \chi_{UV}/\chi_0 = 1.710 \chi$.
Inserting this into equation 12 of \cite{mad93}, the argument of the
logarithm, i.e., $[90(\chi_{UV}/\chi_0)/n] + 1$ can be written as $[53 G_0 /n]
+ 1$, in agreement with equation \ref{eqn:dissociate1} when the latter is
written to accommodate irradiation over $4\pi$ sr.

%\clearpage

\section{RADIO BRIGHTNESS UNIT CONVERSIONS} \label{app:conversions}

It is common practice to report results from imaging synthesis observations of
radio spectral lines as contour diagrams of constant surface brightness using
``J beam$^{-1}$ \kmps'' as the unit.\footnote{Or, quite incorrectly, as
``J \kmps'' (e.g.\ \citet{she02}).  Another interesting (and equally erroneous)
variant is to give single dish spectra in units of ``K beam$^{-1}$'' (e.g.\
Fig.\ 4 in \citet{sch02}).}  This unfortunate combination of units is
telescope dependent through the use of the beam unit in the denominator.  The
conversion to telescope-independent units such as K \kmps\ (which we use here
in eqs.\ \ref{eqn:TmbCO} and \ref{eqn:TmbHI}) requires knowledge of the beam
area of the radio telescope in steradians, which is not well defined for any
radio synthesis imaging telescope owing to the absence of information at short
interferometer spacings.  Fortunately (at least for now), most data processing
systems perform an image restoration step that sets the final point spread
function to be a two-dimensional Gaussian.  This provides a definite (but not
necessarily accurate) value for the two-dimensional integral over the beam,
and thereby ``solves'' the unit conversion problem.  If we take the
FWHMs of this Gaussian to be $\theta_1$ and $\theta_2$, then
the conversion for 21~cm \HI\ image contours may be done with the
relation

\begin{equation}
\frac{T_b(\HI)}{dS/d\Omega} = \frac{6.057 \times 10^5}
{\theta_1 \times \theta_2},
\label{eqn:convertHI}
\end{equation}

\noindent for $T_b$ in K, $dS/d\Omega$ in Jy beam$^{-1}$, and
$\theta_1$ and $\theta_2$ in arcseconds.
The corresponding relation for the 2.6~mm line of CO(1--0) is:

\begin{equation}
\frac{T_b({\rm CO})}{dS/d\Omega} = \frac{91.97}{\theta_1 \times \theta_2},
\label{eqn:convertCO}
\end{equation}

\noindent with the same units as the previous equation.  We emphasize
again that these temperatures are well defined only for completely resolved
sources.  If the telescope beam does not resolve the source, the data can
provide only a lower limit to the true brightness; such lower limits are
usually called ``main beam brightness temperatures'' and labelled $T_{mb}$.

\onecolumn	% for the figures

\clearpage

\clearpage

\clearpage

\clearpage

\begin{thebibliography}{}

\bibitem[Allen et al.(1995)]{all95}
Allen, R.J., Le Bourlot, J., Lequeux, J., Pineau Des For\^ets, G., \&
Roueff, E.~1995, \apj, 444, 157

\bibitem[Allen et al.(1997)]{all97}
Allen, R.J., Knapen, J., Bohlin, R., \& Stecher, T.~1997, \apj, 487, 171

\bibitem[Blitz(1993)]{bli93}
Blitz, L.~1993, in \textit{Protostars \& Planets III}, eds. E.H. Levy
\& J.I. Lunine (Tucson; Univ. Arizona Press), 125

\bibitem[Draine(1978)]{dra78}
Draine, B.T.\ 1978, \apjs, 36, 595

\bibitem[Draine \& Bertoldi(1996)]{dra96}
Draine, B.T., \& Bertoldi, F.\ 1996, \apj, 468, 269

\bibitem[Goldschmidt \& Sternberg(1995)]{gol95}
Goldschmidt, O., \& Sternberg, A.~1995, \apj, 439, 256

\bibitem[Habing(1968)]{hab68}
Habing, H.J.\ 1968, Bull.\ Astr.\ Inst.\ Netherlands, 19, 421

\bibitem[Hollenbach et al.(1991)]{hol91}
Hollenbach, D.J., Takahashi, T., \& Tielens, A.G.G.M.~1991, \apj, 377, 192

\bibitem[Hollenbach \& Tielens(1999)]{hol99}
Hollenbach, D.J., \& Tielens, A.G.G.M.~1999, Revs. Mod. Phys., 71, 173

\bibitem[Kaufman et al.(1999)]{kau99}
Kaufman, M.J., Wolfire, M.G., Hollenbach, D.J., \& Luhman, M.L.~1999,
\apj, 527, 795

\bibitem[Li et al.(2002)]{li02}
Li, W., Evans, N.J., Jaffe, D.T., van Dishoeck, E.F., \& Thi, W.-F.\ 2002,
\apj, 568, 242

\bibitem[Maddalena \& Thaddeus(1985)]{mad85}
Maddalena, R.J., \& Thaddeus, P.~1985, \apj, 294, 231

\bibitem[Madden et al.(1993)]{mad93}
Madden, S.C., Geis, N., Genzel, R., Herrmann, F., Jackson, J., 
Poglitsch, A., Stacey, G.J., \& Townes, C.H.~1993, \apj, 407, 579

\bibitem[Schinnerer \& Scoville(2002)]{sch02}
Schinnerer, E., \& Scoville, N.~2002, \apj, 577, L103

\bibitem[Sheth et al.(2002)]{she02}
Sheth, K., Vogel, S.N., Regan, M.W., Teuben, P., Harris, A.J.,
\& Thornley, M.D.~2002, \aj, 124, 2581

\bibitem[Shull(1978)]{shu78}
Shull, J.M.~1978, \apj, 219, 877

\bibitem[Smith et al.(2000)]{smi00} Smith, D.A., Allen, R.J., Bohlin, R.C.,
Nicholson, N., \& Stecher, T.P.~2000, \apj, 538, 608

\bibitem[Stecher \& Williams(1967)]{ste67}
Stecher, T.P., \& Williams, D.A.~1967, \apj, 149, L29

\bibitem[Sternberg(1988)]{ste88}
Sternberg, A.~1988, \apj, 332, 400

\bibitem[Sternberg \& Dalgarno(1989)]{ste89}
Sternberg, A., \& Dalgarno, A.~1989, \apj, 338, 197

\bibitem[Sternberg \& Dalgarno(1995)]{ste95}
Sternberg, A., \& Dalgarno, A.~1995, \apjs, 99, 565 

\bibitem[Tielens \& Hollenbach(1985)]{tie85}
Tielens, A.G.G.M., \& Hollenbach, D.~1985, \apj, 291, 722

\bibitem[Williams \& Maddalena(1996)]{wil96}
Williams, J.P., \& Maddalena, R.J.~1996, \apj, 464, 247

\bibitem[Witt \& Johnson(1973)]{wit73}
Witt, A.N., \& Johnson, M.W.~1973, \apj, 181, 363 

\bibitem[Wolfire et al.(1990)]{wol90}
Wolfire, M.G., Tielens, A.G.G.M., \& Hollenbach, D.~1990, \apj, 358, 116

\bibitem[Wolfire et al.(1993)]{wol93}
Wolfire, M.G., Hollenbach, D., \& Tielens, A.G.G.M.~1993, \apj, 402, 195

\end{thebibliography}
\end{document}